\documentclass[a4paper,11pt]{article}
\usepackage{pos}
\usepackage{slashed}
\usepackage{hyperref}

\title{Hadron structure via Generalized Parton Distributions}
\ShortTitle{Hadron structure via GPDs}

\author*[a]{Shohini Bhattacharya}

\affiliation[a]{Department of Physics, University of Connecticut, Storrs, CT 06269, USA}

\emailAdd{shohinib@uconn.edu}

\abstract{Recent advancements have made it possible to approximate light-cone correlation functions in lattice QCD by computing their Euclidean counterparts. In these proceedings, we review key developments in this approach and explore their direct implications for Generalized Parton Distributions (GPDs). Furthermore, we emphasize the pivotal role of GPDs in uncovering the internal structure of hadrons and beyond.}

\FullConference{The 41st International Symposium on Lattice Field Theory (LATTICE2024)\\
 28 July - 3 August 2024\\
Liverpool, UK\\}

\begin{document}
\maketitle

\section{Introduction}
\begin{wrapfigure}{r}{5.5cm}  
    \centering
    \includegraphics[width=5.5cm]{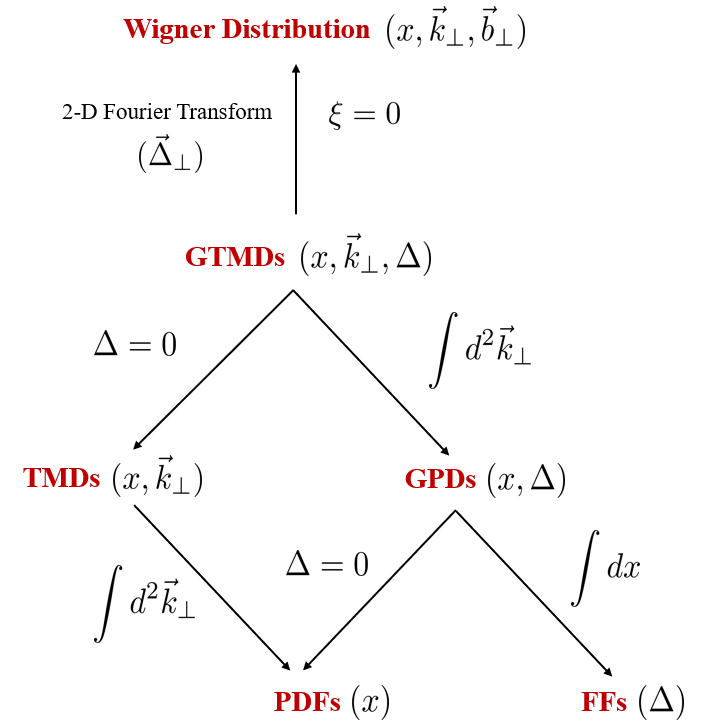}
    \caption{Hierarchy of parton distributions, emphasizing the central role of GPDs in bridging parton distribution functions (PDFs) and form factors (FFs).} 
    \label{fig:GTMD_mother}
\end{wrapfigure}
The non-perturbative structure of nucleons is encoded in various distribution functions, each offering a different perspective on how quarks and gluons are arranged within nucleons. The most fundamental are parton distribution functions (PDFs), which provide a one-dimensional view of how partons carry momentum along the nucleon’s longitudinal direction. Moving beyond this, form factors (FFs) extend the picture to two dimensions, revealing, for instance, how charge distributions vary spatially within the nucleon. However, to fully capture the complexity of parton dynamics, we need even more detailed distributions.

This is where transverse momentum-dependent distributions (TMDs) and generalized parton distributions (GPDs) come into play. TMDs unravel the intricate motion of partons in both longitudinal and transverse momentum space, while GPDs bridge the gap between PDFs and FFs by encoding correlations between momentum and spatial distributions. All the physics embedded in these distributions can be reconstructed from a more general class of functions known as generalized TMDs (GTMDs) or Wigner functions, which provide the most comprehensive description of parton correlations. Mapping these distributions is a central goal of nucleon structure studies, which we aim to achieve with the Electron-Ion Collider (EIC). Currently under construction at Brookhaven National Laboratory, the EIC will be a state-of-the-art facility for exploring quark and gluon dynamics in nucleons with unprecedented precision. Complementing these experimental efforts, lattice QCD provides first-principles calculations that offer crucial insight, particularly for quantities that are difficult to access experimentally.

As shown in Fig.~\ref{fig:GTMD_mother}, the hierarchical connections between these distributions are crucial for developing a complete understanding of nucleon structure. In these proceedings, we focus on GPDs, a powerful tool for nucleon tomography, among other key aspects that we will discuss below.

\section{Generalized Parton Distributions: Definition and Motivation}
We start by recalling the definition of light-cone GPDs that specify 
the state of quarks inside a spin-1/2 nucleon. A light-cone GPD correlator for quarks is 
defined through the Fourier transform of off-forward matrix elements of non-local quark 
fields (see for instance Ref.~\cite{Diehl:2003ny}). The definition reads:
\begin{equation}
F^{[\Gamma]}(x, \Delta; \lambda, \lambda') = \frac{1}{2} \int \frac{dz^{-}}{2\pi} 
 e^{i k \cdot z} \langle p';\lambda'| \bar{\psi} (-\tfrac{z}{2})\, \Gamma 
 \, {\cal W}(-\tfrac{z}{2},\tfrac{z}{2}) \psi (\tfrac{z}{2})|p;\lambda \rangle 
 \bigg |_{z^{+}=0,\vec{z}_{\perp}=\vec{0}_{\perp}} \, .
\label{e: corr_standard_GPD}
\end{equation} 
Color gauge invariance of this non-local quark-quark correlator is enforced by the Wilson line:
\begin{equation}
{\cal W}(-\tfrac{z}{2},\tfrac{z}{2})\bigg |_{z^{+}=0,\vec{z}_{\perp}=\vec{0}_{\perp}} 
= {\cal P} \, {\rm exp} \bigg ( -ig \int^{\tfrac{z^{-}}{2}}_{-\tfrac{z^{-}}{2}} 
 \, dy^{-} A^{+}(0^{+},y^{-},\vec{0}_{\perp}) \bigg ) \,. 
\label{e: wilson_line_standard_GPD}
\end{equation}
In Eq.~(\ref{e: wilson_line_standard_GPD}), $g$ represents the strong coupling constant, 
and $A^{+}$ is the light-cone plus component of the gluonic field. The incoming (outgoing) 
nucleon state in Eq.~(\ref{e: corr_standard_GPD}) is specified by its four-momentum 
$p \, (p')$ and helicity $\lambda \, (\lambda')$. The kinematical variables of interest are given by:
\begin{equation}
P= \frac{1}{2}(p+p'), \qquad
\Delta =p'-p, \qquad 
\xi =\frac{p'^{+}-p^{+}}{p'^{+}+p^{+}}, \qquad
t = \Delta^{2}.
\end{equation}
Here, $P$ represents the average momentum of the nucleon, $\Delta$ quantifies the momentum transfer, $\xi$, the skewness variable, describes the longitudinal momentum transfer, and $t$ is the usual Mandelstam variable that characterizes the squared momentum transfer. Depending on the gamma matrix $\Gamma$ that is sandwiched between the quark fields in Eq.~(\ref{e: corr_standard_GPD}), 
different GPDs emerge, each encoding distinct information about quarks inside nucleons. 
At twist-2, the choices \(\Gamma = \gamma^+, \gamma^+\gamma_5, i \sigma^{j+}\gamma_5\) (where \(j\) is a transverse index) span a total of eight GPDs.
More specifically: Unpolarized quark distributions ($H, E$) are obtained from the vector projection $\gamma^{+}$. Longitudinally polarized quark distributions ($\widetilde{H}, \widetilde{E}$) arise from the axial-vector projection $\gamma^{+}\gamma_{5}$. Transversely polarized quark distributions ($H_{T}, E_{T}, \widetilde{H}_{T}, \widetilde{E}_{T}$) are derived from the tensor projection $i\sigma^{j+}\gamma_{5}$.
A generic GPD is a function of the variables $(x, \xi, t)$, where $x$ denotes the average longitudinal momentum fraction, while $\xi$ and $t$ are the previously defined momentum transfer variables.

There are several compelling reasons to study GPDs. First, for certain kinematic regimes, the Fourier transforms of GPDs can be related to impact parameter distributions, which describe the spatial density of partons inside nucleons in a combined one-dimensional momentum and two-dimensional spatial representation. This is known as nucleon tomography, a major motivation for studying GPDs. Recently, high-resolution lattice calculations of these impact parameter distributions have provided unprecedented insights (see Ref.~\cite{Bhattacharya:2023ays}). At the very least, such studies allow us to examine the differences between up-quark and down-quark distributions inside nucleons, shedding light on their internal structure with remarkable detail.

Second, one can also gain insights into the angular momentum distribution of partons inside nucleons, which is essential for fully understanding the nucleon's spin structure. As noted in Ref.~\cite{Ji:1996ek}, the total angular momentum of quarks, for instance, can be related to specific moments of the GPDs: $J^q =\int_{-1}^{1} dx \, x \left( H^q + E^q \right) |_{t=0}$.

Third, very interestingly, one can also gain insight into the mechanical properties of nucleons. For example, what are the pressure and shear forces experienced by quarks inside nucleons? The fundamental object that addresses these questions is the so-called Energy-Momentum Tensor, which can be parameterized in terms of gravitational form factors (GFFs). These form factors provide a window into the mechanical structure of protons, illustrating how they would interact with gravitons. Of course, in nature, we cannot scatter protons off gravitons in experiments. However, we can indirectly infer their mechanical structure by exploiting the connections between GFFs and GPDs~\cite{Polyakov:2002wz}. GFFs are simply certain moments of GPDs. For instance, we have relations such as $A(t) + \xi^2 D(t) = \int^1_{-1} dx \, x H(x,\xi,t)$, where \( A(t) \) and \( D(t) \) are GFFs. This connection allows us to probe fundamental aspects of nucleon structure that would otherwise be inaccessible.

Lastly, recent studies have uncovered previously unexplored connections between chiral and trace anomalies and specific GPDs, providing a fresh perspective on symmetry breaking and mass generation mechanisms for particles such as the $\eta'$ meson and glueballs. This marks the first time these effects have been associated with GPDs. Specific relations are given by  $
\widetilde{E} \sim \frac{1}{t - m^2_{\eta'}}, \quad H, E \sim \frac{1}{t - m^2_G}$. 
See, for example, Refs.~\cite{Bhattacharya:2023wvy,Tarasov:2025mvn}.

We conclude this section by noting that extracting GPDs through traditional processes, such as Deeply Virtual Compton Scattering and meson production, presents significant challenges, necessitating alternative approaches. These difficulties arise because differential cross-sections are sensitive only to $x$-integrated GPDs rather than the distributions themselves (see, for example, Ref.~\cite{Ji:1996ek}). In this context, lattice QCD has emerged as a crucial tool and is the focus of these proceedings, which summarize key calculations of $x$-dependent GPDs made possible in recent years.

\section{Calculating parton distributions in Lattice QCD: The essence of the quasi-distribution approach}
In this section, we explore the quasi-distribution approach, a Euclidean method for calculating parton distributions from lattice QCD~\cite{Ji:2013dva}. To gain deeper insight, we first calculate the twist-2 (unpolarized) PDF $f_1(x)$ through an elegant perturbative calculation. We focus on a Quark Target Model calculation of a ``ladder-type diagram''. The contribution of this diagram to the light-cone $f_{1}(x)$ is given by:
\begin{align}
f_{1} (x)
&= -\dfrac{i g^{2} C_{F} \mu^{2\epsilon} g_{\mu \nu}}{4} \int^{\infty}_{-\infty} \dfrac{d^{n}k}{(2\pi)^{n}} \dfrac{{\rm Tr} \big [ u \, \overline{u} \, \gamma^{\nu} \, (\slashed{k}+m_{q}) \, \gamma^{+} \, (\slashed{k}+m_{q}) \, \gamma^{\mu} \big ]}{(k^{2} -m^{2}_{q}+ i\varepsilon)^{2} ((p-k)^{2}-m^{2}_{g} + i\varepsilon)} \, \delta \bigg ( x- \dfrac{k^{+}}{p^{+}} \bigg ) \dfrac{1}{p^{+}} \, .
\label{e:f1_matching_step1}
\end{align}
Here, $(u, \bar{u})$ are the Dirac spinors corresponding to the external on-shell quark states, and $(m_{g}, m_{q})$ are the gluon and quark masses. A standard approach to handling the loop integrals in Eq.~(\ref{e:f1_matching_step1}) is to first decompose the loop momentum $k$ into $(k^{+}, k^{-}, k_{\perp})$ components. The $k^{+}$ integral is straightforward due to the delta-function in Eq.~(\ref{e:f1_matching_step1}). The $k^-$ integral can be evaluated using the contour technique, while the $k_{\perp}$-integral exhibits both ultraviolet (UV) and infrared (IR) divergences. We adopt Dimensional Regularization (DR) for the UV divergences, and retain $m_{g} \neq 0$ for the IR regularization. The final result for $f_{1}(x)$ is:
\begin{align}
f_{1} & = \dfrac{\alpha_{s}C_{F}}{2\pi} (1-x) \bigg ( {\cal P}_{UV} + \ln \dfrac{\mu^{2}_{UV}}{x m^{2}_{g}} - 2\bigg ) \, ,
\label{e:f1_matching_step2}
\end{align}
where ${\cal P}_{UV}$ is defined as:
\begin{displaymath}
{\cal P}_{UV} = \frac{1}{\epsilon_{UV}} + \ln 4\pi - \gamma_E \, .
\label{e:def_PUV}
\end{displaymath}
Next, consider performing the same calculation using a purely \textit{spatial correlator}, and subsequently boosting the target to infinite momentum in the $z$-direction. Choosing $\gamma^{3}$ as the operator, the starting expression is:
\begin{align}
f_{1, \rm{Q}} (x)
& = -\dfrac{i g^{2} C_{F} \mu^{2\epsilon} g_{\mu \nu}}{4} \int^{\infty}_{-\infty} \dfrac{d^{n}k}{(2\pi)^{n}} \dfrac{{\rm Tr} \big [ u \, \overline{u} \, \gamma^{\nu} \, (\slashed{k}+m_{q}) \, \gamma^{3} \, (\slashed{k}+m_{q}) \, \gamma^{\mu} \big ]}{(k^{2} -m^{2}_{q}+ i\varepsilon)^{2} ((p-k)^{2}-m^{2}_{g} + i\varepsilon)} \, \delta \bigg ( x- \dfrac{k^{3}}{p^{3}} \bigg ) \dfrac{1}{p^{3}} \, ,
\label{e:f1_matching_step3}
\end{align}
where $p^{3}$ is the $z$-component of the target momentum, and $k^{3}$ represents its fraction. Here, the loop momentum $k$ is decomposed into $(k^{0}, k_{\perp}, k^{3})$. Performing the $k^{3}$-integral using the delta function and evaluating $k^{0}$ via contour techniques, we find that taking the limit $p^{3} \rightarrow \infty$ \textit{before} integrating over $k_{\perp}$ reproduces Eq.~(\ref{e:f1_matching_step2}). (Note that subleading power corrections scale as $1/p_3^2$ and are typically accompanied by nontrivial, $x$-dependent prefactors that exhibit divergences.)\footnote{Here, $p^2_3$ should be understood as $(p^3)^2$. We used this for ease of notation.
}

The analysis above demonstrates that parton physics can alternatively be formulated in terms of Euclidean correlations in the Infinite Momentum Frame (IMF). However, can this approach be replicated on the lattice? Unfortunately, the answer is no, due to the interplay between the two limits, \( (p^3 \to \infty, \int d^2 k_\perp) \). On the lattice, the maximum achievable value of \( p^3 \) is constrained by the finite lattice spacing. In addition to this, the worsening signal-to-noise ratio at higher momenta is a significant limitation. As momentum increases, the statistical noise in lattice QCD calculations grows faster than the signal, making the extraction of meaningful results increasingly difficult. In fact, in practice, this issue imposes a much more stringent constraint, necessitating that lattice practitioners work with finite \( p^3 \). Consequently, one must first perform the transverse momentum integral, \( \int d^2 k_\perp \). This leads to the key question: what happens if we keep \( p^3 \) finite and repeat the above calculation?
More specifically, what is the outcome when one systematically expands in powers of $1/p^2_3$ \textit{after} performing the $k_\perp$ integral?\footnote{More technically, the limitation on $p^3$ makes the UV cutoff $\Lambda$ finite ($a \sim \Lambda$, where $a$ is the lattice spacing). One is therefore forced to address UV renormalization prior to the large-momentum limit in lattice calculations.} The result is given by:
\begin{align}
    f_{1, \rm{Q}} (x; p^{3}) & = \dfrac{\alpha_s C_F}{2\pi}
    \begin{cases}
        (1-x) \ln \dfrac{x}{x - 1} + 1 & x > 1  \\[0.5cm]
        (1-x) \ln \dfrac{4 (1-x)p^{2}_{3}}{m_g^2} + x & 0 < x < 1 \\[0.5cm]
        (1-x) \ln \dfrac{x - 1}{x} - 1 & x < 0 
    \end{cases} \, + \, \mathcal{O}\bigg ( \dfrac{1}{p^2_3} \bigg ) .
    \label{e:f1_matching_step4}
\end{align}
Clearly, the result in Eq.~(\ref{e:f1_matching_step4}) differs significantly from Eq.~(\ref{e:f1_matching_step2}).
 Several key observations emerge:
\begin{itemize}
    \item The quasi-PDF $f_{1, \rm{Q}}$ exhibits finite support outside the ``physical'' region $0 < x < 1$.
    \item IR divergences appear only within the $0 < x < 1$ region. Importantly, the IR pole structures of $f_{1, \rm{Q}}(x)$ and $f_{1}(x)$ are identical, both containing the term $-(1-x) \ln m^{2}_{g}$. Additionally, the quasi-PDF acquires an explicit dependence on $p^3$.
    \item Interestingly, UV divergences, such as those encountered in the light-cone PDF $f_1(x)$, are absent. As a result, the $x$-dependent quasi-PDF $f_{1, \rm{Q}}(x)$ is UV-finite. However, the same UV divergences arising from $\int^{\infty} d^{2}k_{\perp}$ in $f_1(x)$ instead appear in the integral $\int^{\infty}_{-\infty} dx$ for $f_{1, \rm{Q}}(x)$. This feature necessitates an additional renormalization of quasi-PDFs in the ``unphysical'' regions $x > 1$ and $x < 0$.
\end{itemize}
The key takeaway is that the order of these two limits—whether $p^3 \to \infty$ is taken before or after the $k_\perp$-integral—matters. The limits do not commute, leading to two important consequences:
\begin{itemize}
    \item Nontrivial differences in the UV behavior of the quasi-PDF and the light-cone PDF.
    \item No change in the IR singularities between the quasi-PDF and the light-cone PDF—this is the core principle of the quasi-distribution approach.
\end{itemize}
Due to this agreement in the IR, the two distributions can be related through a perturbative procedure called ``matching''. The generic structure of a matching formula is,
\begin{align}
    \tilde{q}(x, \mu, P^3) = \int_{-1}^{1} \frac{dy}{|y|} C\left(\frac{x}{y}, \frac{\mu}{P^3}\right) q(y, \mu) 
+ \mathcal{O} \left( \frac{\Lambda_{\text{QCD}}^2}{(P^3)^2}, \frac{M_N^2}{(P^3)^2} \right) \, ,
\end{align}
where $\tilde{q}$ and $q$ denote the quasi-PDF and light-cone PDF, respectively, \( C \) is a perturbatively calculable matching coefficient, \( \mu \) is the renormalization scale, and \( p^{3} = (x/\xi)P^{3} \), with \( P^{3} \) representing the hadron momentum, \( \Lambda_{\rm QCD} \) the confinement scale, and \( M_N \) the nucleon mass.
 The matching for GPDs has a more intricate structure:
\begin{align}
    \tilde{q}(x, \xi, t, \mu, P^3) = \int_{-1}^{1} \frac{dy}{|y|} C\left(\frac{x}{y}, \frac{\xi}{y}, \frac{\mu}{P^3}\right) q(y, \xi, t, \mu) 
+ \mathcal{O} \left( \frac{\Lambda_{\text{QCD}}^2}{(P^3)^2}, \frac{M_N^2}{(P^3)^2}, \frac{t}{(P^3)^2} \right).
\end{align}
This complexity arises due to the multi-dimensional nature of GPDs. Currently, GPD matching is only known up to one-loop order, with key references including~\cite{Ji:2015qla,Liu:2019urm,Yao:2022vtp}. Note that, for the specific case of $\xi=0$, the GPD matching coincides with the PDF matching, at least at twist-2. For unpolarized PDFs, the state-of-the-art calculation is the two-loop matching~\cite{Li:2020xml}, which represents a significant advancement in the field.

We conclude this section by noting that, in addition to the quasi-distribution approach, several other methods exist. Among these, the pseudo-distribution approach~\cite{Radyushkin:2019owq} has gained particular popularity and has recently been employed to extract GPDs~\cite{Bhattacharya:2024qpp,HadStruc:2024rix}. Another notable method is the Compton amplitude approach~\cite{Hannaford-Gunn:2024aix}, which has been used to compute both moments and the $x$-dependence of GPDs. A comprehensive overview of these Euclidean approaches, along with a detailed report on their corresponding lattice results, can be found in Ref.~\cite{Cichy:2021lih}. In these proceedings, our focus will be on presenting results within the quasi-distribution approach.

\section{Compilation of Lattice QCD results for $x$-dependent GPDs: Progress and Developments}
\subsection{First calculations of twist-2 GPDs}
\begin{figure}[t]
    \centering
    \begin{minipage}{0.44\textwidth}
        \centering
        \includegraphics[width=\textwidth]{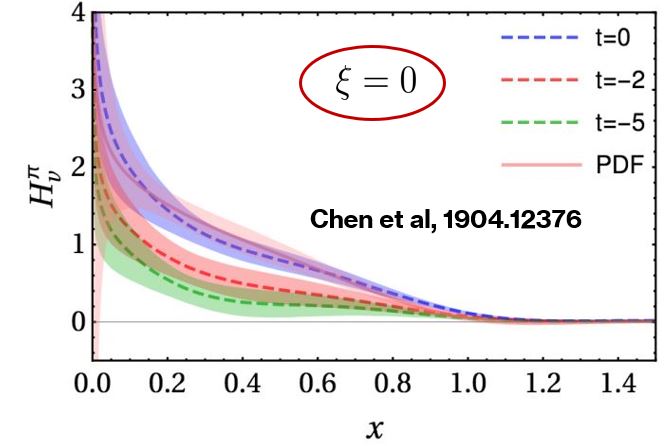}
    \end{minipage}
    \hfill
    \begin{minipage}{0.48\textwidth}
        \centering
        \includegraphics[width=\textwidth]{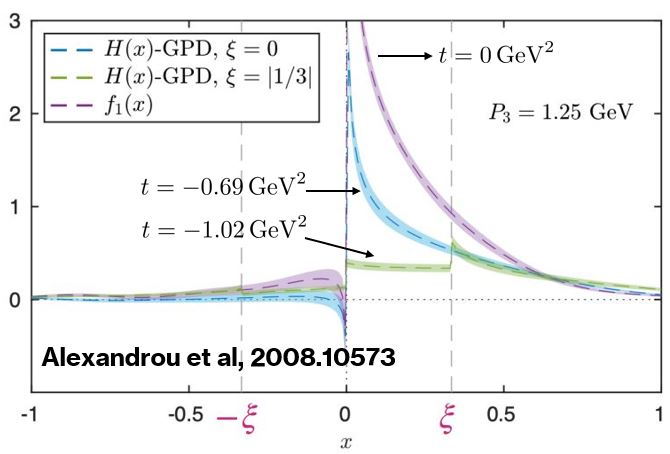}
    \end{minipage}
    \caption{Left: Pion GPD as a function of $x$ for $\xi =0$ at various values of $t$. Right: Proton GPD as a function of $x$ for both $\xi=0$ and $\xi \neq 0$, with different values of $t$.}
    \label{fig:combined1}
\end{figure}
\begin{figure}[t]
    \centering
    \begin{minipage}{0.5\textwidth}
        \centering
        \includegraphics[width=\textwidth]{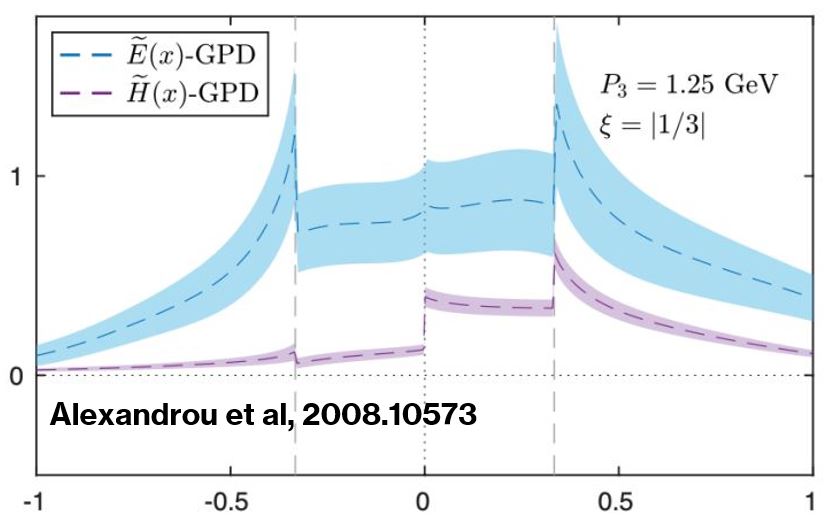}
    \end{minipage}
    \hfill
    \begin{minipage}{0.44\textwidth}
        \centering
        \includegraphics[width=\textwidth]{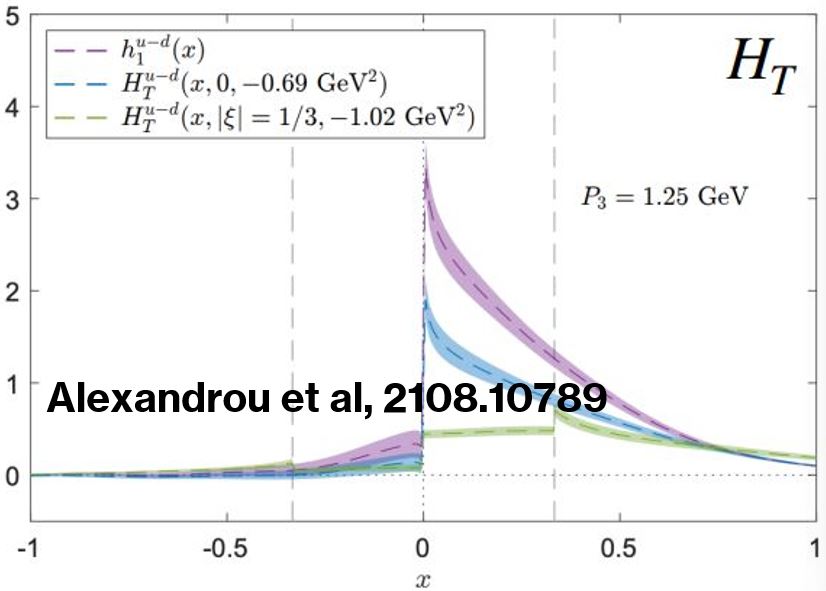}
    \end{minipage}
    \caption{Left: Proton helicity GPDs \( \tilde{H} \) and \( \tilde{E} \) at \( \xi \neq 0 \) as a function of \( x \). Right: Proton transversity GPD \( H_T \) at both \( \xi = 0 \) and \( \xi \neq 0 \) for different values of \( t \) as a function of \( x \).}
    \label{fig:combined2}
\end{figure}
The first-ever calculation of the pion GPD within the quasi-distribution approach was presented in Ref.~\cite{Chen:2019lcm}. The left plot in Fig.~\ref{fig:combined1} shows the pion GPD as a function of $x$ at $\xi=0$ for various values of $t$. Although the uncertainties appear large, the results exhibit the expected qualitative behavior of a GPD—namely, as $t$ increases, the distribution gradually flattens.

The right plot in Fig.~\ref{fig:combined1} presents the first-ever calculation of the proton (unpolarized) GPD as a function of \( x \) at both \( \xi = 0 \) and \( \xi \neq 0 \) for different values of \( t \), as reported in Ref.~\cite{Alexandrou:2020zbe}.
 A few notable qualitative features emerge from the plot, particularly the differences in behavior between the small-\( x \) and large-\( x \) regions. Specifically, there is greater sensitivity to changes in \( t \) at small \( x \), whereas as \( x \to 1 \), this sensitivity to \( t \) completely disappears. This observation aligns with the power-counting analysis suggested in Ref.~\cite{Yuan:2003fs}. Despite large and uncontrolled uncertainties, this calculation successfully predicted the qualitative behavior of the GPD over the entire range of \( x \), which was otherwise inaccessible through experiments. Furthermore, Refs.~\cite{Alexandrou:2020zbe,Alexandrou:2021bbo} extracted all the twist-2 GPDs, some of which are presented in Fig.~\ref{fig:combined2}. These results are particularly remarkable given that, at the time, no GPD extractions from experimental data existed for the full \( x \)-dependence. A recurring question arising from these plots concerns the meaning of the discontinuities at \( x = \pm \xi \). These discontinuities are not physical but rather artifacts of power corrections in GPDs, which diverge at these points.

\subsection{First calculations of twist-3 GPDs}
\begin{figure}[t]
    \centering
    \begin{minipage}{0.44\textwidth}
        \centering
        \includegraphics[width=\textwidth]{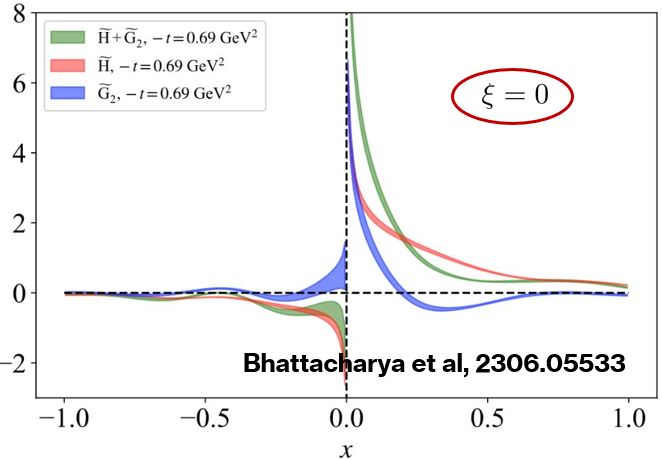}
    \end{minipage}
    \hfill
    \begin{minipage}{0.48\textwidth}
        \centering
        \includegraphics[width=\textwidth]{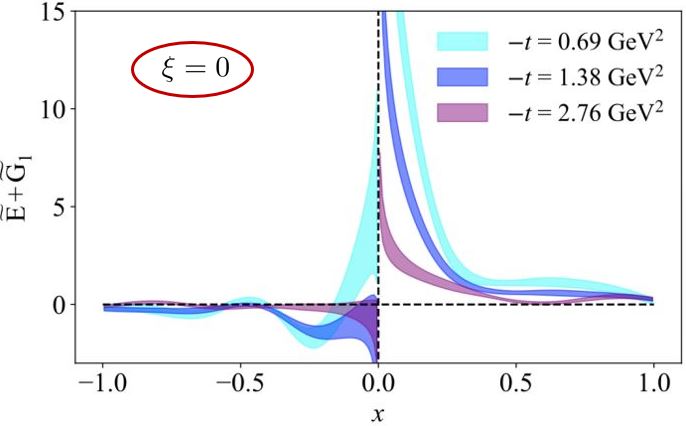}
    \end{minipage}
    \caption{Left: Proton twist-2 GPD \( \tilde{H} \) and twist-3 combination \( \tilde{H} + \tilde{G}_2 \) as functions of \( x \), compared against each other at \( \xi = 0 \). Right: Proton twist-3 GPD combination \( \tilde{E} + \tilde{G}_1 \) at \( \xi = 0 \) as a function of \( x \).}
    \label{fig:combined4}
\end{figure}
Recently, there has been significant progress in the calculation of twist-3 GPDs~\cite{Bhattacharya:2023nmv}. These quantities are essential to study due to their comparable magnitude to twist-2 GPDs and their crucial role in capturing quark-gluon-quark correlations inside hadrons.

Fig.~\ref{fig:combined4} illustrates, on the left, the proton twist-2 GPD \( \tilde{H} \) and the twist-3 combination \( \tilde{H} + \tilde{G}_2 \) as functions of \( x \), compared against each other at \( \xi = 0 \). This comparison clearly demonstrates that twist-2 and twist-3 GPDs can be sizeable. This is an important result because, although twist-3 GPDs appear suppressed by the hard scale of the process in experiments, the GPDs themselves are not. Thus, the ability to compute them on the lattice is highly valuable. On the right of Fig.~\ref{fig:combined4}, we show the twist-3 combination \( \tilde{E} + \tilde{G}_1 \) at \( \xi = 0 \). Since the norm of the twist-3 GPD \( \tilde{G}_1 \) is expected to vanish at \( \xi=0 \), it is reasonable to expect that the majority of the signal in the plot originates from the twist-2 GPD \( \tilde{E} \). Notably, \( \tilde{E} \) is typically inaccessible at \( \xi=0 \) because it drops out from the matrix element. However, here we obtain a glimpse of this otherwise inaccessible function at \( \xi=0 \) through its twist-3 linear combination, which is a significant achievement. Furthermore, by analyzing the $t$-dependence of the GPDs, we found that \( \tilde{E} \) exhibits a pronounced dependence on \( t \) near the origin. This suggests that we are observing the pion-pole dominance picture, $\tilde{E}_u - \tilde{E}_d \sim \frac{1}{t - m_{\pi}^2}$, 
being realized at the level of the GPD for the first time—a phenomenon well known at its corresponding form-factor level, where the \( x \)-integrated version of the GPD corresponds to the form factor, see Fig.~\ref{fig:GTMD_mother}.

These results are remarkable, considering that we have no experimental information on the \( x \)-dependence of GPDs, and even less on higher-twist GPDs. While there are certainly theoretical and lattice-related challenges that need further improvement, these calculations demonstrate the impressive capabilities of lattice QCD in exploring GPD structures beyond what is accessible through experiments.

\subsection{A fresh take: Revisiting GPD calculations in a different kinematic frame}
\begin{figure}[t]
    \centering
    \begin{minipage}{0.48\textwidth}
        \centering
        \includegraphics[width=\textwidth]{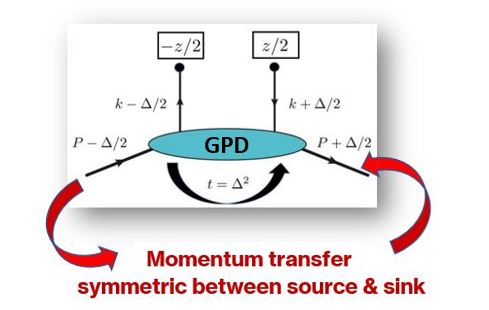}
    \end{minipage}
    \hfill
    \begin{minipage}{0.48\textwidth}
        \centering
        \includegraphics[width=\textwidth]{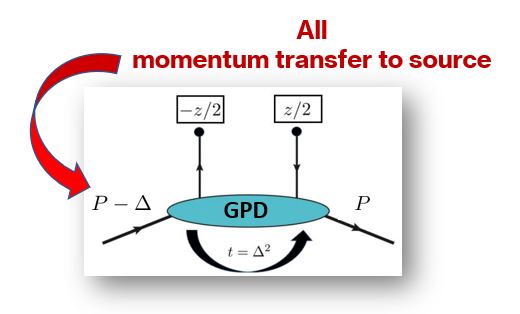}
    \end{minipage}
    \caption{Left: Symmetric frame of reference for calculating GPDs. Note the symmetric appearance of momentum transfer $\Delta$ between incoming and outgoing nucleon states. Right: Asymmetric frame of reference for calculating GPDs. Note that the momentum transfer is only on one of the nucleon states, here the incoming state.
}
    \label{fig:combined5}
\end{figure}
All the GPD results we discussed earlier come with a challenge. Traditionally, these calculations have been performed in the so-called ``symmetric frame'', as shown in the left plot of Fig.~\ref{fig:combined5}. In this frame, the momentum transfer $\Delta$ is symmetrically distributed between the incoming (source) and outgoing (sink) states of the nucleon. However, lattice calculations in these frames are highly inconvenient because every time we change the value of $t$, adjustments are required in both states' momenta, imposing a significant computational burden.

A key way to resolve this practical challenge is to address the question: Can we probe the same physics from a different frame of reference? The answer is yes. If we select a frame where all the momentum transfer is assigned to a single state, such as the incoming state (see the right plot of Fig.~\ref{fig:combined5}), we can circumvent this issue~\cite{Bhattacharya:2022aob,Bhattacharya:2023jsc}. Such a frame is called an ``asymmetric frame''. By performing lattice QCD calculations of GPDs in asymmetric frames, we achieve:
\begin{itemize}
    \item A significant reduction in computational cost.
    \item Access to a broader range of $t$ enables a higher-resolution exploration of the partonic structure of nucleons~\cite{Bhattacharya:2023ays,Bhattacharya:2024wtg}.
\end{itemize}
In Refs.~\cite{Bhattacharya:2022aob,Bhattacharya:2023jsc}, significant theoretical advancements were made for working in an asymmetric frame. First, a Lorentz-covariant formalism was developed to compute quasi-GPDs in \textit{any} frame. Second, quasi-GPDs at finite boost are generally affected by power corrections. Within this framework, it was shown that multiple definitions of quasi-GPDs exist, each differing from the others by power corrections. This non-uniqueness naturally raises questions about which definition converges faster.

The starting point for providing a Lorentz-covariant formulation is the equation below:  
\begin{align} 
F^{\mu} (z,P,\Delta) &= \bar{u}(p',\lambda') \bigg [ \dfrac{P^{\mu}}{m} A_1 + m z^{\mu} A_2 + \dfrac{\Delta^{\mu}}{m} A_3 + i m \sigma^{\mu z} A_4 \nonumber \\
& + \dfrac{i\sigma^{\mu \Delta}}{m} A_5 + \dfrac{P^{\mu} i\sigma^{z \Delta}}{m} A_6 + m z^{\mu} i\sigma^{z \Delta} A_7 + \dfrac{\Delta^{\mu} i\sigma^{z \Delta}}{m} A_8  \bigg ] u(p, \lambda) \, .
\label{e:main}
\end{align}
Here, $F^{\mu} = \langle p', \lambda' | \bar{q} (-z/2) \gamma^\mu q(z/2) | p, \lambda\rangle |_{z=0, \vec{z}_\perp = \vec{0}_\perp}$, and we observe eight Dirac structures, each multiplied by eight linearly independent amplitudes: 
\begin{equation}
A_i \equiv A_i (z\cdot P, z \cdot \Delta, t = \Delta^2, z^2).
\end{equation}  
One may wonder about the significance of this equation. In Eq.~(\ref{e:main}), we have systematically shifted all the kinematic dependence into specific Dirac structures, while all non-perturbative information resides in the amplitudes. This has significant consequences. Before addressing these points, we first note that, for example, unpolarized quasi-GPDs are defined using the operator $\gamma^0$:
\begin{equation}
F^{0} (z, P^s, \Delta^s) = \bar{u}^s(p^{s'}, \lambda') \left[ \gamma^0 H^s_{\text{Q}(0)}(z, P^s, \Delta^s) 
+ \frac{i\sigma^{0\mu} \Delta^s_{\mu}}{2M} E^s_{\text{Q}(0)}(z, P^s, \Delta^s) \right] u^s(p^s, \lambda) \, ,
\label{e:main2}
\end{equation}
where the superscript `\( s \)' on the variables denotes the symmetric frame. By mapping Eqs.~(\ref{e:main}) and~(\ref{e:main2}), we can establish a relation such as  
\begin{equation}
H^s_{\rm Q(0)}(x, \xi, t; P^{3}) = \sum_i c^s_{i} \,  A_i \, .
\end{equation}  
This implies that, for example, the quasi-GPD \( H_{\rm Q(0)} \) in the symmetric frame can be expressed as a linear combination of amplitudes that are independent of the frame in which they are calculated, with \( c^s_{i} \) serving as frame-dependent kinematic prefactors. Consequently, we can construct any quasi-GPD in our preferred symmetric frame using matrix elements computed entirely in the asymmetric frame, significantly reducing computational costs for lattice practitioners.

Regarding the second point mentioned above about power corrections, we present an expression for the light-cone GPD \( H \), which is a Lorentz-invariant quantity:
\begin{equation}
H \left( z \cdot P^{s/a}, z \cdot \Delta^{s/a}, (\Delta^{s/a})^2 \right) = A_1 + \frac{\Delta^{s/a} \cdot z}{P^{s/a} \cdot z} A_3.
\label{e:lc}
\end{equation}
The corresponding quasi-GPD, which is generally a frame-dependent quantity at finite momentum boost, is expressed in the symmetric frame (which we focus on to illustrate this point) and is given by:
\begin{align}
H^{s}_{{\rm Q}(0)}(z, P^s, \Delta^s) & = A_1 + \frac{\Delta^{0,s}}{P^{0,s}} A_3 - \frac{m^2 \Delta^{0,s} z^3}{2 P^{0,s} P^{3,s}} A_4 +
\left[ \frac{(\Delta^{0,s})^2 z^3}{2 P^{3,s}} - \frac{\Delta^{0,s} \Delta^{3,s} z^3 P^{0,s}}{2 (P^{3,s})^2} - \frac{z^3 (\Delta^s_{\perp})^2}{2 P^{3,s}} \right] A_6 \nonumber \\
& + \left[ \frac{(\Delta^{0,s})^3 z^3}{2 P^{0,s} P^{3,s}} - \frac{(\Delta^{0,s})^2 \Delta^{3,s} z^3}{2 (P^{3,s})^2} - \frac{\Delta^{0,s} z^3 (\Delta^s_{\perp})^2}{2 P^{0,s} P^{3,s}} \right] A_8.
\label{e:quasi}
\end{align}
By comparing Eqs.~(\ref{e:lc}) and~(\ref{e:quasi}), we observe that there are significantly more amplitudes at the level of quasi-GPDs than in the light-cone GPD case. This suggests that, at least naively, one must reach higher and higher momentum values to suppress these ``additional contaminations'' and approach the light-cone GPD limit. However, this is not a practical feature due to the limitation imposed by finite lattice spacing, which restricts the maximum achievable momentum. The key question we addressed in our work is whether there exists a definition of quasi-GPDs in which, \textit{a priori}, all these additional amplitudes are absent.

\begin{figure}[t]
    \centering
    \begin{minipage}{0.55\textwidth}
        \centering
        \includegraphics[width=\textwidth]{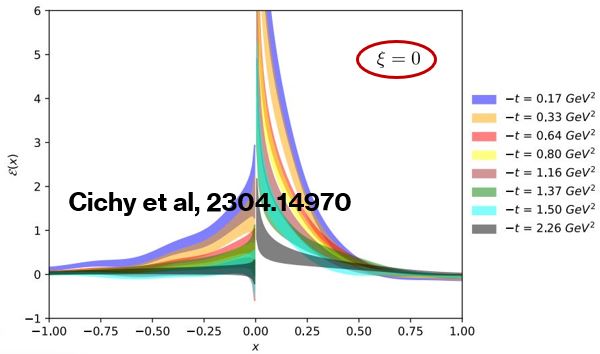}
    \end{minipage}
    \hfill
    \begin{minipage}{0.44\textwidth}
        \centering
        \includegraphics[width=\textwidth]{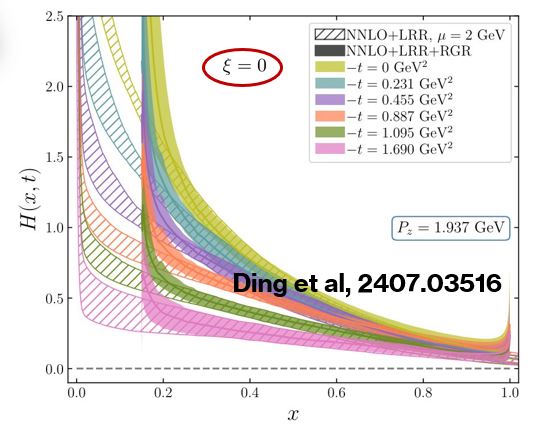}
    \end{minipage}
    \caption{Left: The unpolarized proton GPD $E$, derived within the amplitude formalism from an asymmetric frame using the Lorentz-invariant definition. We present this specific example of the $E$ GPD because the Lorentz-invariant definition yields a more precise result compared to the traditional operator $\gamma^0$. Right: The unpolarized pion GPD $H$, obtained within the same amplitude formalism and asymmetric frame.}
    \label{fig:combined6}
\end{figure}
The main idea we propose is that the answer to the question raised above is yes—it is possible to construct a quasi-GPD in which all extra amplitudes are absent \emph{a priori} by appropriately modifying the operator definition of the quasi-GPD. The schematic structure is given by:
\begin{equation}
    {\mathcal{H}} \to c_0 \langle \bar{\psi} \gamma^0 \psi \rangle + c_1 \langle \bar{\psi} \gamma^1 \psi \rangle + c_2 \langle \bar{\psi} \gamma^2 \psi \rangle .
    \label{e:li}
\end{equation}
The schematic structure demonstrates that a quasi-GPD can be expressed as a sum of bilinear quark operators, \( \gamma^{1,2} \), weighted by certain frame-dependent kinematic factors and added to the usual \( \gamma^0 \). These additional \( \gamma^{1,2} \) operators play a crucial role in canceling extra amplitudes that would otherwise appear in the \( \gamma^0 \)-only formulation. This ensures that the resulting quasi-GPDs retain the same functional form as their light-cone counterparts, thereby preserving Lorentz invariance even at finite momentum boost. The schematic structure of the Lorentz-invariant definition of quasi-GPD is given by:
\begin{equation}
\mathcal{H} \left( z \cdot P^{s/a}, z \cdot \Delta^{s/a}, (\Delta^{s/a})^2, z^2 \right)
= A_1 + \frac{\Delta^{s/a} \cdot z}{P^{s/a} \cdot z} A_3 .
\end{equation}
This expression has the same functional form as the light-cone GPD in Eq.~(\ref{e:lc}). The key difference, however, is that in this case, the amplitudes implicitly depend on power corrections, i.e., \( A_i \equiv A_{i}(z^2 \neq 0) \), whereas for light-cone GPDs, they are evaluated at \( z^2=0 \). Consequently, one might expect these Lorentz-invariant quasi-GPD definitions to converge more rapidly to light-cone GPDs. However, this reasoning is likely too simplistic and requires further substantiation. It is possible that the presence of additional amplitudes (explicit power corrections) interacts nontrivially with implicit power corrections, leading to an overall reduction in power corrections for the traditional operator definition compared to the Lorentz-invariant one. As a result, the actual convergence of different quasi-GPD definitions may, in reality, be governed by the underlying dynamics. Empirically, we have observed that the absence of certain amplitudes does not necessarily guarantee faster convergence—while \( E \) shows significant improvement, \( H \) is not substantially affected. For \( \tilde{H} \), the traditional operator definition exhibits better convergence.

We present specific GPDs for protons and pions using the amplitude formalism and asymmetric frame formulation in Fig.~\ref{fig:combined6} from Refs.~\cite{Cichy:2023dgk,Ding:2024hkz}. Notably, the spanned values of $t$ highlight a key advantage we previously discussed—working in asymmetric frames allows access to a broader range of $t$ values. This advancement is now achievable within lattice calculations.

\section{Summary}
In these proceedings, we have highlighted the significant recent progress in studying the partonic structure of hadrons using lattice QCD through Euclidean correlators. Lattice QCD calculations are particularly impactful in areas where experimental access is challenging—GPDs being a prime example. Among the various Euclidean approaches, the quasi-distribution approach has seen remarkable advancements, making it the central focus of our discussion.

Over the years, major strides have been made in computing both leading-twist and higher-twist GPDs for pions and protons, particularly in determining their $x$-dependence—an aspect that remains elusive in experiments. As the field evolves, efforts are underway to reformulate these calculations using computationally more efficient and faster techniques, a development we have explored here.

Beyond methodological improvements, recent work has also focused on integrating lattice-QCD results with experimental data for specific processes within the GPD framework, allowing for a more detailed extraction of nucleon tomography and the total angular momentum carried by valence quarks~\cite{Cichy:2024afd}. This marks an important step toward bridging lattice calculations with phenomenology. Naturally, as new experimental facilities such as the EIC become operational, the experimental landscape will evolve as well. This opens the door for a truly complementary relationship between lattice QCD and phenomenology, where lattice predictions can help interpret experimental results while experimental constraints refine lattice methodologies.

We are in an exciting era of rapid progress, with more breakthroughs on the horizon. GPDs play a crucial role in unraveling the intricate dynamics of partons inside nucleons, and advancements in lattice QCD calculations will provide invaluable insights into the fundamental structure of matter—insights that will be particularly significant in the EIC era.

\acknowledgments I sincerely thank the organizers of the Lattice 2024 Conference for inviting me to give a plenary talk. I am also deeply grateful to Krzysztof Cichy for reviewing these proceedings and providing valuable feedback.


\end{document}